# Temporal extensivity of Tsallis' entropy and the bound on entropy production rate

Sumiyoshi Abe* and Yutaka Nakada

*Institute of Physics, University of Tsukuba, Ibaraki 305-8571, Japan*

**Abstract**  The Tsallis entropy, which is a generalization of the Boltzmann-Gibbs entropy, plays a central role in nonextensive statistical mechanics of complex systems. A lot of efforts have recently been made on establishing a dynamical foundation for the Tsallis entropy. They are primarily concerned with nonlinear dynamical systems at the edge of chaos. Here, it is shown by generalizing a formulation of thermostatistics based on time averages recently proposed by Carati [A. Carati, Physica A **348**, 110 (2005)] that, whenever relevant, the Tsallis entropy indexed by $q$ is temporally extensive: linear growth in time, i.e., finite entropy production rate. Then, the universal bound on the entropy production rate is shown to be $1/|1-q|$. The property of the associated probabilistic process, i.e., the sojourn time distribution, determining randomness of motion in phase space is also analyzed.

―――――――――――――――――――

*Corresponding author. E-mail address: suabe@sf6.so-net.ne.jp

## I. INTRODUCTION

The Boltzmann-Gibbs entropy and ordinary statistical mechanics based on it require a system under consideration to be in the strongly chaotic regime. It may be possible to view this point in Boltzmann's *Stosszahlansatz* (i.e., molecular chaos hypothesis), which, combined with the *H*-theorem, enables one to obtain the thermal equilibrium distribution in the long-time limit. That is, there exists no correlation between colliding particles. Over a century after Boltzmann's period, physicists and mathematicians have made a lot of efforts to build a bridge between statistical mechanics and microscopic dynamics. As known today, one way to satisfy the above requirement is to demand that the system possesses at least one positive Lyapunov exponent [1,2].

The Kolmogorov-Sinai entropy [1,2], which is the nonlinear dynamics counterpart of the Boltzmann-Gibbs entropy, quantifies the strength of dynamical chaos. A point of crucial importance is its change due to dynamical evolution: it grows linearly in time, that is, the entropy production rate is constant. Then, the Pesin identity [1,2] tells that the rate is given by the sum of positive Lyapunov exponents. In analogy between statistical mechanics and chaotic dynamics, this linear growth of the Kolmogorov-Sinai entropy corresponds to thermodynamic extensivity of the Boltzmann-Gibbs entropy [1], i.e., linear scaling with respect to the number of particles. And, it is considered that extensivity of entropy is an indispensable requirement, with which thermodynamics can be constructed. This may be the case even if system energy is nonextensive [3].

It should be noted that, in general, it is sufficient to realize the linear growth of the Kolmogorov-Sinai entropy and thermodynamic extensivity of the Boltzmann-Gibbs entropy only in the long-time limit and the thermodynamic limit, respectively.

A physical quantity is said to be *temporally extensive* if it grows linearly in time. Thus, for example, the Kolmogorov-Sinai entropy possesses temporal extensivity for chaotic dynamical systems.

Now, the situation becomes different for complex dynamical systems prepared at the edge of chaos, in which their maximum Lyapunov exponents strictly vanish. In this case, the standard Pesin identity becomes trivial ($0=0$), offering no information, and the Kolmogorov-Sinai entropy fails to be temporally extensive. Then, the so-called generalized Lyapunov exponents or, $q$-Lyapunov exponents [4-10] and the Tsallis entropy [11] may become physically relevant. The authors of Refs. [4-10] have studied several kinds of nonlinear dynamical systems at the edge of chaos and have found that the generalized Pesin identity with the $q$-Lyapunov exponents holds and the Tsallis entropy with the entropic index $q$ different from unity (i.e., different from the Kolmogorov-Sinai limit) is temporally extensive. Most of these works are numerical, but remarkably Ref. [8] rigorously shows these results for the logistic map by using exact renormalization group analysis.

Generalized statistical mechanics based on the Tsallis entropy, termed nonextensive statistical mechanics [12], is considered to be a consistent and unified framework for describing complex statistical systems and is currently under vital investigation.

Yet, there is another approach to thermostatistics, which is based on time averages. This method proposed by Carati [13,14] allows us to establish a connection between a probabilistic process in a phase space and definition of entropy. Since it directly treats dynamics as a time series, it may shed new light on a long-standing problem of statistics and dynamics.

In this paper, we show temporal extensivity of the Tsallis entropy, without recourse to employing any specific dynamical system. This is executed by generalizing the above-mentioned formulation of thermostatistics based on time averages. Then, we derive the universal bound on the Tsallis-entropy production rate. We shall also see what probabilistic process in a phase space underlies nonextensive statistical mechanics.

The paper is organized as follows. In Sec. II, the formulation of thermostatistics based on time averages presented in Refs. [13,14] is reviewed with some modifications. In Sec. III, such a formulation is generalized, and then it is shown that the Tsallis entropy is temporally extensive. In Sec. IV, the universal bound on the Tsallis-entropy production rate is presented. Analysis of the probabilistic process associated with the Tsallis entropy and nonextensive statistical mechanics is performed in Sec. V. Finally, Sec. VI is devoted to concluding remarks.

Throughout this paper, the Boltzmann constant, $k_B$, is set equal to unity for the sake of simplicity.

## II. THERMOSTATISTICS BASED ON TIMES AVERAGES

In this section, we wish to review Carati's formulation of thermostatistics based on time averages, which is equivalent to Boltzmann-Gibbs theory. But, at the same time, we shall make some modifications of it.

Consider a system in a phase space $M$. Its dynamics yields a sequence $\{x_n\}_{n=1,2,\cdots}$. This sequence is generated by a map, $\phi: M \to M$, defining the iteration $x_{n+1} = \phi(x_n)$. The average of a certain physical quantity $A$ over a fixed long time interval $1 \leq n \leq N$ ($N \gg 1$) is given by

$$\overline{A}(x_0) \equiv \frac{1}{N} \sum_{n=1}^{N} A(x_n). \tag{1}$$

This quantity is random, depending on the initial data, $x_0$.

A procedure of coarse graining is to divide $M$ into a lot of cells: $L_1, L_2, \cdots, L_K$ ($K \gg 1$). Let $A_i$ and $n_i$ be the representative value of $A$ in the $i$th cell $L_i$ and the number of times the system visits $L_i$, respectively. Then, Eq. (1) is well approximated as follows:

$$\overline{A}(x_0) \cong \sum_{i=1}^{K} A_i \frac{n_i}{N}. \tag{2}$$

In this representation, $\{n_i\}_{i=1,2,\cdots,K}$ is random, depending on $x_0$.

Next, let $P(n_i \leq n) \equiv F(n_i)$ be the cumulative probability that the cell $L_i$ is visited $n_i \, (\leq n)$ times by the system. It essentially describes the sojourn time characterizing system dynamics in the phase space [15]. Then, the average value of $A$ to be compared with observation is given by

$$<\bar{A}> = \frac{1}{N} \sum_{i=1}^{K} A_i <n_i>, \tag{3}$$

where

$$<n_i> = \frac{\int_{\Sigma n_j = N} \prod_{j=1}^{K} dF(n_j) \, n_i}{\int_{\Sigma n_j = N} \prod_{j=1}^{K} dF(n_j)}, \tag{4}$$

provided that the integral should be understood as the Lebesgue-Stieljes integral. Eq. (3) can be rewritten as

$$<\bar{A}> = -\frac{1}{N} \frac{\partial}{\partial \lambda} \ln Z(\lambda) \bigg|_{\lambda=0} \tag{5}$$

with the generating function

$$Z(\lambda) = \int_{\Sigma n_j = N} \prod_{j=1}^{K} dF(n_j) \, e^{-\lambda \sum_i A_i n_i}$$

$$= \int \prod_{j=1}^{K} dF(n_j) \, e^{-\lambda \sum_i A_i n_i} \delta(N - \Sigma_i n_i). \tag{6}$$

If the constraint is imposed on the energy

$$U = \frac{1}{N}\sum_{i=1}^{K} \varepsilon_i n_i \qquad (7)$$

with $\varepsilon_i$ the representative value of the system energy in the cell $L_i$, Eq. (6) should be replaced by

$$Z_U(\lambda) = \int \prod_{j=1}^{K} dF(n_j)\, e^{-\lambda \sum_i A_i n_i} \delta(N - \Sigma_i n_i)\, \delta(U - \Sigma_i \varepsilon_i n_i / N)$$

$$= \frac{N}{(2\pi)^2} \int_{-\infty}^{\infty} dk_1 \int_{-\infty}^{\infty} dk_2 \int \prod_{j=1}^{K} dF(n_j)\, e^{ik_1 UN + ik_2 N}$$

$$\times \prod_{i=1}^{K} e^{-n_i(\lambda A_i + ik_1 \varepsilon_i + ik_2)}. \qquad (8)$$

Accordingly, the average value in Eq. (5) is modified as

$$<\bar{A}>_U = -\frac{1}{N}\frac{\partial}{\partial \lambda}\ln Z_U(\lambda)\bigg|_{\lambda=0}. \qquad (9)$$

It is convenient to define the function, $\chi(\zeta)$, as follows:

$$\int dF(n)\, e^{-n\zeta} = e^{\chi(\zeta)}. \qquad (10)$$

Then, Eq. (8) becomes

$$Z_U(\lambda) = \frac{N}{(2\pi)^2} \int_{-\infty}^{\infty} dk_1 \int_{-\infty}^{\infty} dk_2 \ e^{ik_1 UN + ik_2 N + \sum_i \chi(\lambda A_i + ik_1 \varepsilon_i + ik_2)}. \tag{11}$$

In the large-$N$ limit, the integrals are evaluated by the method of steepest descents. The steepest-descent condition leads to

$$U = -\frac{1}{N} \sum_{i=1}^{K} \varepsilon_i \ \chi'(ik_1 \varepsilon_i + ik_2), \tag{12}$$

$$N = -\sum_{i=1}^{K} \chi'(ik_1 \varepsilon_i + ik_2), \tag{13}$$

in the limit $\lambda \to 0$, where $\chi'$ is the derivative of $\chi$ with respect to the argument. As can be seen from Eq. (9), the relation

$$<\overline{A}>_U = -\frac{1}{N} \sum_{i=1}^{K} A_i \ \chi'(ik_1 \varepsilon_i + ik_2) \tag{14}$$

holds in the leading order of $N$. (It is discussed in Ref. [13] how fluctuations around the steepest descents are small.) Eqs. (12)-(14) imply that $-\chi'(\theta \varepsilon_i + \alpha)$ is the average number of times that the cell $L_i$ is visited by the system, where the analytic continuations, $ik_1 = \theta$ and $ik_2 = \alpha$, have been made. Therefore, defining the sojourn time probability

$$p_i = -\frac{\chi'(\theta \varepsilon_i + \alpha)}{N}, \tag{15}$$

we have

$$<\bar{A}>_U = \sum_{i=1}^{K} A_i \, p_i. \tag{16}$$

An interesting discovery of Carati is that Boltzmann-Gibbs statistical mechanics can be reproduced if $F(n)$ is Poissonian, i.e., the completely random process. In this case, Eq. (10) becomes

$$\sum_{n=0}^{\infty} e^{-Np} \frac{(Np)^n}{n!} e^{-n\zeta} = e^{Npe^{-\zeta} - Np}, \tag{17}$$

showing

$$\chi(\zeta) = Npe^{-\zeta} - Np, \tag{18}$$

where $p$ is a positive constant to be determined later.

In anticipation of the relation between $\chi(\zeta)$ and the free energy, consider the following Legendre transformation:

$$s(v_i) = v_i \zeta_i + \chi_0(\zeta_i), \tag{19}$$

$$v_i = -\chi'_0(\zeta_i), \tag{20}$$

with

$$\zeta_i = \theta \varepsilon_i + \alpha. \tag{21}$$

In these equations, we are using only the relevant part of $\chi(\zeta_i)$ in Eq. (18)

$$\chi_0(\zeta_i) = N p e^{-\zeta_i}, \tag{22}$$

since the term, $-Np$, is nothing but a constant shift of the free energy. From Eq. (18), $\zeta_i$ and $v_i$ are calculated to be

$$\zeta_i = -\ln(p_i / p), \tag{23}$$

$$v_i = \chi_0(\zeta_i) = N p e^{-\zeta_i} = N p_i, \tag{24}$$

where $p_i$ is given in Eq. (15). Then, the entropy is found to be given by the Boltzmann-Gibbs entropy

$$S = \sum_{i=1}^{K} s(v_i) = \sum_{i=1}^{K} [v_i \zeta_i + \chi_0(\zeta_i)]$$

$$= -N \sum_{i=1}^{K} p_i \ln p_i, \tag{25}$$

provided that $p$ has been fixed as

$$p = \frac{1}{e}. \tag{26}$$

The thermodynamic entropy, $S_{th}$, may be defined by

$$S_{th} = \frac{S}{N}. \tag{27}$$

From Eqs. (12), (20), and (27), it follows that

$$\frac{\partial S_{th}}{\partial U} = \theta. \tag{28}$$

Thus, $\theta$ is found to be the inverse temperature, $\theta = 1/T \equiv \beta$ ($k_B \equiv 1$). Now, using Eqs. (15) and (22), and recalling $N = -\sum_i \chi'(\theta \varepsilon_i + \alpha)$, one obtains

$$p_i = \frac{e^{-\beta \varepsilon_i}}{Z(\beta)}, \qquad Z(\beta) = \sum_{i=1}^{K} e^{-\beta \varepsilon_i}, \tag{29}$$

which is precisely the canonical distribution in Boltzmann-Gibbs statistical mechanics.

Note that $S$ in Eq. (25) is temporally extensive, linearly scaling with respect to $N$, and the thermodynamic entropy in Eq. (27) is actually the entropy production rate.

### III. TEMPORAL EXTENSIVITY OF TSALLIS ENTROY

We wish to construct the Tsallis entropy [11] and nonextensive statistical mechanics in the spirit of the discussion in the preceding section. Although this issue has already been addressed in Ref. [13], several points remain to be clarified and some generalizations are needed. We shall fully develop nonextensive statistical mechanics based on time averages.

First of all, we note that, as shown in Ref. [16], what has to be used in nonextensive statistical mechanics is not the ordinary expectation value in Eq. (16) but the normalized $q$-expectation value

$$<\bar{A}>_{U,q} = \sum_{i=1}^{K} A_i P_i^{(q)}, \tag{30}$$

where $P_i^{(q)}$ is the escort distribution [1] associated with the original distribution, $p_i$:

$$P_i^{(q)} = \frac{(p_i)^q}{\sum_j (p_j)^q}. \tag{31}$$

Since *the time average should be identified with the normalized q-expectation value*, Eq. (15) has to be replaced by

$$P_i^{(q)} = -\frac{\chi'(\theta \varepsilon_i + \alpha)}{N}. \tag{32}$$

However, the Legendre transformation should remain form invariant as in Eqs. (19) and (20). That is, the thermodynamic formalism is kept unchanged.

Now, examine the following deformation of the exponential factor in Eq. (18):

$$\chi(\zeta) = N r_q e_q(-\zeta) - N r_q. \tag{33}$$

In this equation, $e_q(x)$ stands for the $q$-exponential function

$$e_q(x) = [1 + (1-q)x]_+^{1/(1-q)}, \tag{34}$$

with the notation

$$[a]_+ \equiv \max\{0, a\}. \tag{35}$$

In the deformation-free limit $q \to 1$, this function converges to the ordinary exponential function, $e^x$. So, Eq. (33) tends to Eq. (18), if

$$r_q \xrightarrow{q \to 1} p = \frac{1}{e}. \tag{36}$$

As in the preceding section, we can ignore the constant shift, $-Nr_q$, in Eq. (33) and take only the relevant part

$$\chi_0(\zeta) = N r_q e_q(-\zeta). \tag{37}$$

Accordingly, we have

$$v_i = -\chi'_0(\zeta)\big|_{\zeta=\zeta_i=\theta\varepsilon_i+\alpha} = N r_q [e_q(-\zeta_i)]^q$$

$$= N \frac{(p_i)^q}{c_q}, \tag{38}$$

where

$$c_q = \sum_{i=1}^{K} (p_i)^q. \tag{39}$$

$\varepsilon_i$ here is the value of the energy in the cell $L_i$ of a nonextensive complex system, which different from an ordinary simple system considered in the preceding section.

From Eqs. (37) and (38), we have

$$\zeta_i = -\frac{1}{1-q}\left[\left(\frac{v_i}{Nr_q}\right)^{1/q-1} - 1\right], \tag{40}$$

$$\chi_0(\zeta_i) = Nr_q \frac{p_i}{(c_q r_q)^{1/q}}. \tag{41}$$

Using Eqs. (38)-(41), we obtain the Tsallis entropy [11]

$$S_q = \sum_{i=1}^{K} s(v_i) = \sum_{i=1}^{K} [v_i \zeta_i + \chi_0(\zeta_i)]$$

$$= \frac{N}{1-q}\left[\sum_{i=1}^{K} (p_i)^q - 1\right], \tag{42}$$

provided that $r_q$ is chosen to be

$$r_q = \left[ \frac{q}{(c_q)^{1/q}(2-c_q)} \right]^{q/(1-q)}. \tag{43}$$

$r_q$ in fact converges to $p = 1/e$ in Eq. (26) in the limit $q \to 1$, as required in Eq. (36). And, at this level, the entropic index, $q$, is taken to be positive. Clearly, Eq. (42) becomes Eq. (25) in the limit $q \to 1$.

We note that the solution in Eq. (43), and therefore the Tsallis entropy in Eq. (42), can exist if and only if

$$c_q < 2. \tag{44}$$

The marginal case, $c_q \to 2-0$, corresponds to divergently large $r_q$. The consistency condition in Eq. (44) leads to an important result, which will be discussed in the next section.

Thus, we conclude that the Tsallis entropy *necessarily* has temporal extensivity if the time average is given in terms of the escort distribution in Eq. (31) with

$$p_i = \frac{1}{Z_q(\beta)} e_q(-(\theta \varepsilon_i + \alpha)), \qquad Z_q(\beta) = \sum_{i=1}^{K} e_q(-(\theta \varepsilon_i + \alpha)). \tag{45}$$

Finally, we point out that, as in Eqs. (27) and (28),

$$\frac{\partial S_{\text{th}, q}}{\partial U} = \theta \tag{46}$$

holds for $U = \sum_i \varepsilon_i \nu_i / N$ and

$$S_{\text{th},q} = \frac{S_q}{N}, \qquad (47)$$

implying that $\theta$ is the inverse temperature, $1/T$. Therefore, Eq. (44) is precisely the $q$-exponential distribution in nonextensive statistical mechanics [12].

Closing this section, we wish to emphasize that temporal extensivity of the Tsallis entropy plays a crucial role in Eq. (46). Were it not temporally extensive, could temperature not be defined. In this respect, we recall that, also in recent papers [14,17], a discussion has been developed about necessity of extensivity of the Tsallis entropy for temperature to be definable.

## IV. BOUND ON TSALLIS-ENTROPY PRODUCTION RATE

As noted in Sec. III, the Tsallis entropy and associated nonextensive statistical mechanics can consistently be constructed within the time-average formulation if and only if Eq. (44) is satisfied.

The consistency condition in Eq. (44) is always satisfied when $q > 1$, since, in this case, $c_q \equiv \sum_i (p_i)^q < 1$. This immediately gives rise to the bound on the Tsallis-entropy production rate

$$\frac{S_q}{N} < \frac{1}{q-1} \qquad (q>1). \tag{48}$$

The situation becomes nontrivial in the case when $0 < q < 1$. The maximum value of $c_q$ is realized by the equiprobability, $p_i = 1/K$ ($i = 1, 2, \cdots, K$), giving $c_q^{\max} = K^{1-q}$. Therefore, the consistency condition in Eq. (44) yields

$$1 - \frac{\ln 2}{\ln K} < q < 1. \tag{49}$$

Since $K$ is very large, this essentially means the Boltzmann-Gibbs limit, $q \to 1$. However, it is not a physical state considered in nonextensive statistical mechanics of complex systems. The phase space of a complex systems at the edge of chaos has a highly nontrivial structure: only a small number of the cells are visited by the system. Therefore, the equiprobability is not realized. This leads to the conclusion that the phase-space configuration is realized in such a way that Eq. (44) is satisfied for each *system-specific* value of $q \in (0, 1)$. Thus, we find that the Tsallis-entropy production rate has the bound

$$\frac{S_q}{N} < \frac{1}{1-q} \qquad (0 < q < 1). \tag{50}$$

From Eqs. (48) and (50), we conclude that the Tsallis-entropy production rate, which is actually the thermodynamic entropy, has the following universal bound:

$$\frac{S_q}{N} < \frac{1}{|1-q|}. \tag{51}$$

It is of interest to examine this bound for an analytic examples known in the literature. In Ref. [8], the logistic map, $x_{t+1} = 1 - \mu x_t^2$, with the initial condition, $x_0 = 0$, is analytically discussed at the edge of chaos $\mu = 1.4011\cdots$. [This system is dissipative, and therefore the generalized canonical distribution in Eq. (45) is irrelevant and only the property of the Tsallis entropy may be mentioned.] The value of the entropic index of such a system is calculated to be $q = 0.2445\cdots$. On the other hand, the Tsallis-entropy production rate, which is the $q$-Lyapunov exponent, $\lambda_q$, is found to be $\lambda_q = 1/(1-q) = 1.3236\cdots$, This is a marginal case of Eq. (51). In Ref. [9], the conservative triangle map on a torus at the edge of chaos is numerically analyzed with an ensemble of various initial conditions, and the values of $q$ and $\lambda_q$ are found to be about zero and unity, respectively. Again, it is a marginal case.

## V. PROBABILISTIC PROCESS FOR NONEXTENSIVE STATISTICAL MECHANICS

In Sec. III, we have seen that the deformation of $\chi(\zeta)$ from Eq. (18) to Eq. (33) *uniquely* leads to the Tsallis entropy and nonextensive statistical mechanics. (It is in fact unique since the discussion is based on the Legendre transformation and the thermodynamic formalism.) In the case of Eq. (18), the underlying probabilistic process in the phase space is Poissonian, as in Eq. (17). What is the corresponding one for $\chi(\zeta)$ in Eq. (33)? This is the issue to be addressed in this section.

Let $f(n)$ be a distribution function of a discrete random variable, $n = 0, 1, 2, \cdots$. Then, Eq. (10) reads

$$\sum_{n=0}^{\infty} f(n) z^{-n} = \tilde{f}(z), \tag{52}$$

where

$$\tilde{f}(z) = e^{\chi(\zeta)}, \tag{53}$$

$$z = e^{\zeta}. \tag{54}$$

Therefore, $\tilde{f}(n)$ is the $z$-transformation [18] of $f(n)$.

The inverse $z$-transformation is given by [18]

$$f(n) = \frac{1}{2\pi i} \oint_C dz\, \tilde{f}(z)\, z^{n-1}, \tag{55}$$

where $C$ is a circle centered at $z = 0$ (in the complex $z$-plane) surrounding all the poles of $\tilde{f}(n)$.

Upon applying the above inversion formula to Eq. (33), an analytic expression is needed for $\chi(\zeta)$, since $z$ (and therefore $\zeta$) is complex. Accordingly, we have to employ the following expression:

$$\chi(\zeta) = N r_q [1 - (1-q)\zeta]^{1/(1-q)} - N r_q, \tag{56}$$

that is, the previous real expression, $[1-(1-q)\zeta]_+^{1/(1-q)}$, is prolonged to complex $[1-(1-q)\zeta]^{1/(1-q)}$ with a branch point at $\zeta = 1/(1-q)$, in general. Using the change of the variable in Eq. (54), we have

$$f(n) = \frac{1}{2\pi i} \int_{a-i\pi}^{a+i\pi} d\zeta\, e^{\chi(\zeta)}\, e^{n\zeta}, \tag{57}$$

where *a* is the logarithm of the radius of the circle *C*.

Unfortunately, it does not seem to be possible to calculate Eq. (57) analytically in terms of known functions. Here, we make the asymptotic evaluation of $f(n)$ for large values of *n* by the method of steepest descents. Let us rewrite Eq. (57) with Eq. (56) as follows:

$$f(n) = \frac{e^{-N r_q}}{2\pi i} \int_{a-i\pi}^{a+i\pi} d\zeta\, e^{\varphi(\zeta)}, \tag{58}$$

$$\varphi(\zeta) = N r_q [1-(1-q)\zeta]^{1/(1-q)} + n\zeta. \tag{59}$$

Together with the steepest-descent condition, $\varphi'(\zeta_0) = 0$, this approximation yields

$$f(n) \cong \frac{e^{-N r_q + \varphi(\zeta_0)}}{\sqrt{2\pi \varphi''(\zeta_0)}}, \tag{60}$$

where

$$\varphi(\zeta_0) = \frac{1}{1-q}\left[n - qNr_q\left(\frac{n}{Nr_q}\right)^{1/q}\right],\tag{61}$$

$$\varphi''(\zeta_0) = qNr_q\left(\frac{n}{Nr_q}\right)^{2-1/q}\quad (>0).\tag{62}$$

Therefore, for large values of $n$, $f(n)$ asymptotically behaves as follows:

$$f(n) \sim \begin{cases} n^{1/(2q)-1}\exp\left[-\dfrac{qNr_q}{1-q}\left(\dfrac{n}{Nr_q}\right)^{1/q}\right] & (0<q<1) \\ n^{1/(2q)-1}\exp\left(-\dfrac{n}{q-1}\right) & (q>1) \end{cases}.\tag{63}$$

In particular, Eq. (60) tends to the asymptotic form of the Poisson distribution in the limit $q \to 1$

$$f(n)\xrightarrow{q\to 1}\frac{e^{-Np}}{\sqrt{2\pi n}}\exp\left(n - n\ln\frac{n}{Np}\right)\tag{64}$$

with $r_q \to p = 1/e$ ($q \to 1$) in Eq. (36).

## VI. CONCLUDING REMARKS

We have shown by generalizing the formulation of thermostatistics based on time averages that the Tsallis entropy is temporally extensive. We have also shown that, as Boltzmann-Gibbs statistical mechanics, nonextensive statistical mechanics can also be consistently constructed based on time averages. Then, we have presented the universal bound on the Tsallis-entropy production rate. In addition, we have analyzed the probabilistic process (i.e., the sojourn time distribution in the cells) in the phase space and have determined its asymptotic property.

In the present work, we have been concerned with temporal extensivity of entropy. It is considered [1] that temporal extensivity in a fully chaotic dynamical system corresponds to thermodynamic extensivity (i.e., linear scaling with respect to the number of particles) in a statistical system. From the combined viewpoints of statistics and dynamics, extensivity of entropy seems to be an indispensable premise for temperature to be definable [3,14,17]. Recently, it has been shown [19] that the Tsallis entropy behaves as an extensive quantity if a system contains strong correlation of a specific type. It is necessary to understand such correlation from dynamics at the edge of chaos. On the other hand, upon proving temporal extensivity of the Tsallis entropy, we have used the probabilistic process in the phase space. Thus, it is equally necessary to clarify if dynamics at the edge of chaos can certainly yield such a process (and, simultaneously, nonextensive statistical mechanics).

There is yet another interesting issue to be addressed. Strictly speaking, it is sufficient to realize temporal extensivity of entropy only in the large-$N$ limit, i.e., the long-time limit. In recent works [20,21], it has numerically been shown that not only the Tsallis entropy but also the quantum-group entropy [22] and the $\kappa$-entropy [23] possess temporal extensivity in the long-time limit for several dynamical systems at the edge of chaos. These generalized entropies are concave and Lesche-stable [24,25], as the Tsallis entropy is [26]. These facts indicate that there may be a certain level of diversity in microscopic description of thermostatistics of complex systems. It is desirable to be possible to formulate theories for those entropies based on time averages and to show them temporally extensive, as done in the present work. These remaining issues are to be clarified and solved for deeper understanding of statistical mechanics of complex systems.